\documentclass{iaus}
\usepackage{graphicx}

\newcommand{\beq}{\begin{equation}}
\newcommand{\eeq}{\end{equation}}

\newcommand{\kms}{\mbox{ km s$^{-1}$}~}

\newcommand{\Mo}{\mbox{M$_{\odot}$}~}

\title[Dynamical PN MHD Evolution] 
{Dynamical PN Evolution with Magnetic Fields}

\author[Garc\'{\i}a-Segura]   
{Guillermo Garc\'{\i}a-Segura }

\affiliation{Instituto de Astronom\'{\i}a, Universidad Nacional
Aut\'onoma de M\'exico, Apdo Postal 877, Ensenada, Baja California,
Mexico}

\pubyear{2006}
\volume{234}  
\pagerange{1--8}
\date{?? and in revised form ??}
\jname{Planetary Nebulae in our Galaxy and Beyond}
\editors{A.C. Editor, B.D. Editor \& C.E. Editor, eds.}
\begin{document}

\maketitle

\begin{abstract}

Hydrodynamical simulations played an important role in understanding
the dynamics and shaping of planetary nebulae in the past century.
However, hydrodynamical simulations were just a first order approach.
The new millennium arrived with the generalized understanding that 
the effects of magnetic fields were necessary to study the dynamics
of planetary nebulae. Thus, B-fields introduced a whole new number of physical
possibilities for the modeling.
In this paper, we review observational works done in the last 5 years and
several works on magnetohydrodynamics about proto-planetary nebulae, since 
all the effort has been focused on that stage, and discuss different 
scenarios for the origin of magnetized winds, and the relation 
binary-bipolararity. 

\keywords{MHD---Stars: AGB and Post-AGB---Planetary Nebulae: general}

\end{abstract}

\section{Introduction}

The origin and evolution of proto-planetary nebulae (PPNs)  and planetary
nebula (PNs) represent one of the key questions in
our understanding of stellar physics. Modeling the fascinating features
displayed by these objects requires not only a better knowledge of stellar
structure at the AGB stage (and beyond) but also a proper consideration of the
driving mechanisms for mass ejection. The transition from AGB to Post-AGB to PN
central stars involves drastically different conditions at every stage. Whereas
radiation pressure on dust grains is the most likely mechanism at the AGB phase,
as are line-driven winds in the case of PN central stars, for Post-AGB stars the
details of the driving force has been relatively unexplored.

To begin this review, it is interesting to mention the increasing number of 
articles related to magnetic fields in PNs during the 
last years. Before the IAU Symp. 180 (Groningen 1996), the average number of
paper per year was 0.41 (base line of 39 years). For the next 5 years until
the IAU Symp. 209 (Canberra 2001), this number increased up to 6.2.
In the last 5 years, this numbers growths up to 11.8 at the present IAU Symp.
234 (Hawaii 2006). What is even more interesting is that, in the last five 
years, 75 \% of all the papers were observations. Finally, {\bf measurements of
magnetic field intensities and their orientations have been observed !!}. 
This fact will be important in the next five years, since theoretical 
models have something robust to start with.

All the theoretical work done in PNs up to 2001 were focus on the 
line-driven wind theory including weak or moderate magnetic fields frozen 
in the winds (see review by Garc\'{\i}a-Segura 2003).
However, since the paper by Bujarrabal et al. (2001), in which it is stated 
that PPNs could not be explained by radiation 
forces on the winds, most of the work done in the last five years is foccus
on PPNs and the winds from post-AGBs stars. 

In this paper, we first make a short review of the observations done in 
the last five years, and then proceed with the MHD work.

\section{Magnetic field measurements}

\subsection{\rm SiO, ${\rm H}_2$O and OH masers: Zeeman}

15 objects with circular polarized masers due to Zeeman effects have been 
detected so far: S Per (Vlemmings et al. 2001, 2002, 2005); K 3-35
(Miranda et al. 2001, G\'omez et al. 2002, 2005); RT Vir, RCrT, W Hya
(Szymczak et al. 2001); VY CMa, NML CyG, U Her (Vlemmings et al. 2002, 2005);
W43 A (Imai et al. 2002, Vlemmings et al. 2006); OH 17.7-2.0 (Bains et al.
2003, Szymczak et al. 2004); U Ori, Vx SgrI (Vlemmings et al. 2005);
IRAS 20406+2953 (Bains et al.2004); IRAS 07331+0021, IRAS 18266-1239
(Szymczak \& G\'erard 2004). 
The measurements for the magnetic fields intensities goes  from
$10^{-2}$ gauss for OH masers up to several tens of gauss for SiO masers
(Vlemmings et al. 2005, 2006). The most exiting object is W43A, which
present a precesing, magnetically collimated jet (Vlemmings et al. 2006), 
showing also how the field is oriented in the flow. These measurements 
will constrain future MHD models of winds in late AGB stars, and
disk-wind models from interacting binaries and common envelope phases.

\subsection{Nebulae with toroidal fields: dust grain alignments}

The alignment of dust grains under the influence of a magnetic field
can be observed in linear polarization in the submillimetre regime.
The next nebulae show toroidal field configurations: 
NGC 7027 and CRL 2688  (Greaves 2002, Sabin et al. this volume); 
NGC 6537 and NGC 6302 (Sabin et al. this volume).

\subsection{Central stars of PNs: Zeeman}

Magnetic fields of the order of kilogauss have been finally detected
in central stars of PNs. The detected objects are NGC 1360, EGB 5, 
LSS 1362 and Abell 36 (Jordan et al. 2004, 2005). Those central stars
are hot and evolved, thus, these measurements will constrain MHD models
with line-driven, magnetized winds.

\subsection{Large scale fields: Faraday rotation in PNs}

The discovery of a Faraday screen feature associated with
a known astronomical object, the PN S216 have been reported by
Uyaniker (2004). This detection is extremely interesting for two 
reasons, the first one is that it is the first Faraday screen in which the 
distance is known (d $\sim$ 80 pc), and the second one is that
prove the existence of large scale magnetic fields associated with a PN
(S216 is 3 $\times$ 3 pc, $1^{\circ}$ in the sky).

\section{The PPN-wind problem: radiation ruled out}

Winds from AGB stars are thought to be driven by
radiation pressure on dust grains (see review by  Habing 1996), although an
alternative physical mechanism has been proposed by Pascoli (1997) based in
magnetic pressure that is transported out from the stellar interior to the
stellar surface. On the other hand, it is widely accepted that planetary
nebulae (PNs) are powered by line-driven winds emerging from their central
stars, and they are formed from a two-wind dynamic interaction (i.e., Kwok,
Purton \& Fitzgerald 1978). Evidences for this scenario are the large number 
of P-cygni line profiles detected in their central objects (Perinotto 1983).

Post-AGB stars with their associated Proto-planetary nebulae (PPNs) are 
short-lived transition objects between AGB stars and white dwarfs. Their wind
energy source, although unclear, has been usually assumed to be radiation
pressure. However, recent observations of PPNs (Alcolea et al. 2001; Bujarrabal
et al. 2001 and references therein) have revealed that the linear momenta and
kinetic energies associated with these objects are in excess to what can be
provided by radiation pressure alone, in some cases by up to three orders of
magnitude. These large amounts of momentum and energy, as discussed in detail 
by Bujarrabal et al.(2001), cannot be accounted for by either radiation pressure
on dust grains, line-driven winds or continuum-driven winds.

\section{Magnetic-driven winds: a possible solution}

Four scenarios with different types of wind driving mechanisms 
have been suggested in the literature:

\begin{description}

\item{\bf A. Accretion disk wind solutions}

The theory of magneto-centrifugal launched winds from accretion disk, 
where toroidal magnetic fields become finally dominant (e.g. Contopoulos 1995) 
is applied in this scenario. 
Several solutions differs in the accretion mode:
accretion of the primary wind onto the secondary (e.g. Morris 1987, 
Mastrodemos \& Morris 1998); Roche lobe overflow (e.g. Livio, Salzman, 
\& Shaviv 1979; Livio \& Soker 1988);
accretion in the primary after common envelope evolution,
in this case, the disk is formed in the primary star.
Recent works on this scenario are Reyes-Ruiz \& L\'opez (1998),
Blackman et al.(2001b), Frank \& Blackman 2004, and Frank (this volumen).

\item{\bf B. Stellar (Post-AGB) solutions}

The results discussed by Pascoli (1997), based on surface magnetic pressure as
the main driver of the large mass-loss rates in AGB stars, are an
alternative to generate the required mechanical power in the winds of Post-AGB
stars (Garc\'{\i}a-Segura et al. 2005), provided that the generation of
magnetic fields can be efficient in post-AGB stars, as suggested by
Blackman et al. (2001).
There is not yet a clear model of how a single star can achieve this.
One plausible scheme is that the rotation rate and the field strength at the
stellar core increase during the formation of the white dwarf. Thus, the inner
magnetic field becomes stronger as the core contracts and becomes exposed at
the stellar surface when the envelope is peeled-off during the PN formation.
Thus, a strong and dominant toroidal component develops at the interface
between the core and the envelope, where some dynamo
action is expected and which may be responsible for launching a magnetic-driven
wind. Actually, Blackman et al. (2001), Matt et al. (2004) and
Miyaji et al. (this volume) have proposed that
the post-AGB wind is produced by magneto-centrifugal processes when dynamo
activity increases the internal field (see Blackman 2004), and the AGB star
sheds its outer layers, exposing the rotating and magnetized core.
Obviously, more detailed stellar interior studies with rotation and $B$-fields
are needed to
understand the details of this issue. In addition, some authors have suggested
that this may also occur in binary systems and, for instance, Soker (1997)
proposed that the Post-AGB stellar core can be spun-up by a secondary,
increasing the shear between the core and the envelope.

\item{\bf C. Stellar + disk solutions}

This scenario combines the two solutions above, for the case
in which the disk is formed in the primary star
(Blackman et al. 2001, Frank this volume).

\item{\bf D. Dynamo during common envelope phase}

The spiral-in process of a secondary star, or a giant planet, may also be 
able to produce a large shear in the stellar envelope, and raise the 
magnetic field strength by
dynamo activity (Tout \& Reg\"os 2003). These cases link the large mass-loss
rates in the post-AGB stages with a common envelope phase. Unfortunately, there
are very few detailed studies of post-common envelope systems (e.g. Exter,
Pollaco \& Bell 2003).

\end{description}

\section{Magneto-centrifugal core-envelope launching}

A promising avenue, using dynamo amplification at these late evolutionary 
stages, has been discussed by Matt, Frank \& Blackman (2004). 
They use a simplified model in which the
interface between the (rotating and magnetized) stellar core and envelope stores
a large amount of magnetic energy due to the twisting of an originally poloidal
magnetic field. The magnetic energy is extracted from the stellar rotational
energy, causing a rapid spin-down of the proto white dwarf, and is able to
drive a strong and short outburst (this is somehow similar to the "magnetic
bubble" mechanism proposed by Draine (1983), to generate molecular outflows in
star-forming clouds). The outflow can expel the envelope and is termed
"magnetic explosion" by Matt et al. (See also the work by Miyaji et al. in
this volume).

\section{Proto-Planetary Nebula MHD Models}

Garc\'{\i}a-Segura et al. (2005) have followed the magnetic-driven wind 
expansion and nebula formation for six models.
Three of them, models A, B and C have a spherically symmetric initial atmosphere
(their Figure 1), while models D, E and F (their Figure 2) have an 
equatorial density enhancement. 

The numerical solutions show that collimation is well established at the very
early phases of evolution, creating jet-like outflows at locations close to
the star. The inclusion of the density enhancement
(Figure 2) produces, as expected, a narrow equatorial waist without any apparent
direct impact at the polar regions. The polar expansion velocities are similar
for all models with the same input magnetic field; models A and D have $v_{exp}
\sim 30  \kms$, models B and E have $v_{exp} \sim 150 \kms$, while models C and
F have $v_{exp} \sim 390 \kms$. 

As a comparative example, their Figure 3 compares the result of one of the 
models with 1 gauss (model E) at 1,000 yr, with two well known, extremely 
collimated PPNs, He 3-401 (Sahai 2002) and M 2-9 (Schwarz et al.1997). 
It is apparent in this figure that the solution is able to reproduce 
convincingly the extreme collimated shapes, along with the sizes and 
kinematics of these nebulae. 
Similar results also have been computed by Washimi et al. (this volume).

We now turn our attention to the kinetic energy and linear momentum contained
in the outflows from these models. Bujarrabal et al. (2001), as mentioned
earlier, have pointed out that radiation pressure is insufficient to provide the
observed mechanical power in the outflows of PPNs. Figure 6 in
Garc\'{\i}a-Segura et al. (2005) gives the results
for three different values of the surface magnetic fields covering the initial
1000 years of evolution. The data of PPNs from Bujarrabal et al. (2001) are
indicated as crosses in these plots. The values for most of these objects seen
to be well bracketed by models B (1 G) and C (5 G). Therefore, magnetic-driven
winds are able to provide the necessary energy budget to power the outflows of
PPNs.

\section{Magnetic Cycles}

Magnetic cycles, and their associated field reversals, have been proposed
as a plausible origin to the existence of multiple, regularly spaced, and
faint concentric shells around some planetary nebulae observed with the
Space Telescope (Soker 2000; Garc\'{\i}a-Segura et al.2001). In fact, OH
maser observations by
Szymczak et al. (2001) suggest that changes in the polarized maser emission
in some stars could be caused by turbulence in the circumstellar magnetic field
and by global magnetic field reversals. 

Garc\'{\i}a-Segura et al.(2005) explored the effects
of magnetic field reversals in magnetic-driven winds, and compared the
results with objects displaying collimated outflows with periodic outburst
features. An interesting example is He 2-90, a PPN whose symmetric and highly
collimated, knotty, bipolar outflow was described by Sahai \& Nyman (2000).
The radial velocities of the knots have been measured by Guerrero et al. (2001),
and the corresponding proper motions subsequently derived by Sahai et al.
(2002). An interesting, and puzzling characteristic in this case is that the
collimated outflow, or jet, maintains a nearly constant apparent width
throughout
all its extent, i.e. it does not fan out at large distances from the star, and
the velocity of the regularly spaced knots seems to be the same. The ''jet''
speed is somewhere between 150-360 $\kms$, its dynamical time is at least 1400
yr, and the knots are created at the rate of one pair roughly every 35-45 yr.
An extensive numerical study can be found in Lee \& Sahai (2004),
which concluded that the inclusion of a magnetic field was necessary.

\section{Bipolar and elliptical nebulae: the role of binaries} 

The morphology and galactic distribution of PNs was discussed by
Garc\'{\i}a-Segura et al. (2002). We here review the important points.

During the last decade, two important surveys were carried out on both
hemispheres. For the southern hemisphere, {\em The ESO Survey}, the
study was published in a series of papers by Schwarz, Corradi \&
Melnick (1992), Stanghellini, Corradi \& Schwarz (1993) (250 PNs),
Corradi \& Schwarz (1995), and Corradi (2000) (400 PNs). For the
northern hemisphere, {\em The IAC Survey}, the study was published by
Manchado et al.(1996) (243 PNs) and Manchado et al.(2000) (255 PNs).
The first detailed study of the differences between ellipticals and
bipolars was done by Corradi \& Schwarz (1995) with the
ESO Survey (see also Corradi 2000). They found that the bipolar class
has a smaller scale height, 130 pc, than the one for ellipticals, 320
pc. Also, bipolars have the hottest central stars among PNe, and
display smaller deviations from pure circular Galactic rotation than
other morphological types. In addition, bipolars also display the
largest physical dimensions and have expansion velocities of up to an
order of magnitude above the typical values for PNe. These properties,
together with the chemical abundance results by Calvet \& Peimbert (1983),
indicate that bipolar PNe are produced by more massive progenitors
than the remaining morphological classes.

\begin{table}[!b]
\setlength{\tabcolsep}{2em}
\begin{center}
\caption{}
\label{}
\begin{tabular}{lc}
\hline
\hline
Morphological Class &  $< z >$    \\
 (IAC Survey)       &    pc       \\
\hline  
B    &  110 \\
BPS  &  248 \\
E    &  308 \\
EPS  &  310 \\
R    &  753 \\
\hline
\hline
\end{tabular}
\end{center}
\end{table}

The average scale height over the plane, $< z >$, for different PN
morphological classes can be compared with those of different stellar
masses and populations. For instance, we know that massive stars are
located much closer to the galactic plane than the population of stars
with lower initial mass. The outcome from the ESO and IAC surveys is
certainly coincident: the bipolar (B) class has $< z >$ = 130 pc (ESO)
and $< z >$ = 179 pc (IAC), for ellipticals (E) $< z >$ = 320 pc (ESO)
and $< z >$ = 308 pc (IAC), and for rounds (R) $< z >$ = 753 pc (IAC).

In the recent analysis of the IAC Survey by Manchado et al. (2000),
the bipolar and elliptical objects with point-symmetric
features (BPS and EPS) were separated from those which do not present 
such  kind of symmetries, i.e., from the B and E classes respectively.
The new results from the IAC Survey are given in the Table 1.

Comparing the results of the IAC Survey with those described by Miller \&
Scalo (1979) for the average scale heigth of stars with different
masses, an average value equal or smaller than 110 pc corresponds to
stars with initial masses above 1.9 \Mo. Lower mass stars have average
scale heights well above this value. These results are in line with the
pioneering suggestion of Calvet \& Peimbert (1983) and the more recent
discussion made by Garc\'{\i}a-Segura et al.(1999).

The relation between morphology and galactic distribution
was explained in Garc\'{\i}a-Segura et al. (2002) as :

\begin{description}

\item $\bullet$ Bipolars Type I Peimbert = (B):

Small $< z >$  (110 pc) $\Longleftrightarrow$ Massive Progenitor
$\Longleftrightarrow$ Stellar Rotation $\Longleftrightarrow$
$\Omega$ Limit $\Longleftrightarrow$ Classical Bipolarity

\item $\bullet$ Bipolars with Point-Symmetry = (BPS):

Moderate $< z >$  (248 pc) $\Longleftrightarrow$ Non-Massive 
Progenitor in Tidally interacting Binary System $\Longleftrightarrow$ 
Tidal Spin-Up $\Longleftrightarrow$
Shaping by $\Omega$ Limit + MHD Effects + Precession/Wobbling
 $\Longleftrightarrow$ Bipolarity with Point-Symmetry (Lobes, FLIERS, Jets)

\item $\bullet$ Ellipticals = (E):

Medium $< z >$  (308 pc) $\Longleftrightarrow$ Non-Massive Progenitor
$\Longleftrightarrow$ Shaping by MHD Effects $\Longleftrightarrow$
FLIERS \& Jets with Axisymmetry

\item $\bullet$ Ellipticals with Point-Symmetry = (EPS):

Medium $< z >$  (310 pc)  $\Longleftrightarrow$ Non-Massive Progenitor in
Wide Binary System $\Longleftrightarrow$ Shaping by MHD Effects + Precession
$\Longleftrightarrow$ FLIERS \& Jets with Point-Symmetry

\item $\bullet$ Round = (R):

Large $< z >$  (753 pc) $\Longleftrightarrow$ Low-mass Progenitor
$\Longleftrightarrow$ Neither Rotation nor MHD effects

\end{description}

To conclude, the direct signature of a binary is well traced in
nebulae with any kind of point-symmetry (BPS and EPS classes), however, 
the direct link
between binary and bipolar cannot be concluded (the EPS class  is the 
counter example) from the galactic
distribution. It is clear that bipolars are associated with more
massive stars. {\bf If binaries play a role in the formation of  
bipolars, at least one or both stars in the binary system must be massive}.

\begin{acknowledgments}

G.G.-S. thanks partial support by CONACyT grant 43121.

\end{acknowledgments}

\end{document}